\documentclass[showpacs,amsmath,amssymb,twocolumn,floatfix,pra]{revtex4}
\newcommand{\newc}{\newcommand}
\newc{\beq}    {\begin{equation}}
\newc{\eeq}    {\end{equation}}
\newc{\beqa}    {\begin{eqnarray}}
\newc{\eeqa}    {\end{eqnarray}}
\newc{\no}    {\\ \nonumber}

\usepackage[dvips]{graphicx}
\input{epsf}
\begin{document}
\title{Quantum chaos algorithms and dissipative
decoherence with quantum trajectories}
\author{Jae Weon Lee, and Dima L. Shepelyansky}    
\affiliation{Laboratoire de Physique Th\'eorique, UMR 5152 du CNRS,
  Universit\'e Paul Sabatier, 31062 Toulouse Cedex 4, France}
\date{January  21, 2005}

\begin{abstract}
 Using the methods of quantum trajectories
 we investigate the effects of dissipative decoherence in a
 quantum computer algorithm simulating dynamics in various regimes 
 of quantum chaos including dynamical localization,
 quantum ergodic regime and quasi-integrable motion.
 As an example we use the quantum sawtooth algorithm which
 can be implemented in a polynomial number of quantum gates.
 It is shown that the fidelity of quantum computation
 decays exponentially with time
 and that the decay rate is proportional to the number of qubits,
 number of quantum gates and per gate dissipation rate
 induced by external decoherence.
 In the limit of strong dissipation the quantum algorithm
 generates a quantum attractor which may have complex or simple structure.
 We also compare the effects of dissipative decoherence with 
 the effects of static imperfections.
\end{abstract} 

\maketitle

The main fundamental obstacles in realization of quantum computer 
\cite{chuang00} are external decoherence and internal
imperfections. The decoherence is produced by
couplings between the quantum computer and the external
world (see e.g. review \cite{zurek}). 
The internal imperfections appear due to static 
one-qubit energy shifts and
residual couplings between qubits
which exist inside the isolated quantum computer.
These imperfections  may lead to emergence
of quantum chaos and melting of quantum computer
eigenstates \cite{georgeot00,dlsnobel}.
The effects of unitary quantum errors
produced by decoherence and imperfections
on the accuracy of quantum algorithms have been studied 
by different groups using numerical modeling of
quantum computers performing quantum algorithms with
about $10 - 20$ qubits.
The noisy errors in quantum gates produced by external
decoherence are analyzed in
\cite{cirac,paz1,paz2,song1,acat,qcat,braun,terraneo1,song2,levi1,bettelli1,bettelli2} while the errors induced by internal static imperfections
are considered in 
\cite{benenti01,terraneo2,benenti02,pomeransky,frahm,zhirov,levi2}.
The analytical treatment \cite{frahm}
based on the random matrix theory
allows to compare the accuracy bounds for these
two types of errors for quantum algorithms
simulating complex quantum dynamics. 

In fact, a convenient frame for 
investigation of quantum errors effects in quantum computations is given by
models of quantum chaos \cite{chirikov}. Such models describe a
quantum dynamics which is chaotic in the classical limit
and which has a number of
nontrivial properties including dynamical localization of chaos,
quantum ergodicity and mixing in phase space (see e.g. \cite{chirikov}).
It has been shown that for many of such models
the quantum computers with $n_q$ qubits
can simulate the quantum evolution of an exponentially large
state (e.g. with  $N=2^{n_q}$ levels) in a polynomial
number of elementary quantum gates $n_g$ (e.g. with  $n_g=O(n_q^2)$
or $n_g=O(n_q^3)$). The quantum algorithms are now available for
the quantum baker map \cite{schack}, the kicked rotator \cite{kr},
the quantum sawtooth \cite{benenti01,benenti02}
and tent \cite{frahm} maps, the kicked wavelet rotator \cite{terraneo2},
the quantum double-well map \cite{qcat}.
Their further generalization and
development gave quantum algorithms for
the Anderson metal-insulator transition \cite{pomeransky},
electrons on a lattice in a magnetic field and the kicked
Harper model \cite{levi2}.
The quantum algorithm for the quantum baker map has been
implemented experimentally with a NMR based quantum computer \cite{cory}.
 
However till now the quantum chaos algorithms
have been used only for investigation of unitary errors effects.
This is always true for internal static imperfections
but the external decoherence generally leads also to
dissipative errors. The first step in the analysis of 
dissipative decoherence in quantum algorithms
has been done in \cite{carlo1}
on a relatively simple example of entanglement teleportation
along a quantum register (chain of qubits). 
After that this approach has been applied to study
the fidelity decay in the 
quantum baker map algorithm \cite{carlo2}.
In \cite{carlo1,carlo2} the decoherence is investigated in
the Markovian assumption using the master equation
for the density matrix written in the Lindblad form
\cite{lindblad}. Already with $n_q=10 - 20$ qubits 
in the Hilbert space of size $N=2^n_q$ the numerical
solution of the exact master equation becomes enormously
complicated due to a large number of variables in the density
matrix which is equal to $N^2$. Therefore the only possibility 
for numerical studies at large $n_q$ is to use the method
of quantum trajectories for which
the number of variables is reduced to $N$
with additional averaging over many trajectories. 
This quantum Monte Carlo type method appeared as a result
of investigations of open dissipative quantum systems
mainly within the field of quantum optics
but also in the quantum measurement theory 
(see the original works \cite{qt1,qt2,qt3,qt4}).
More recent developments in this field
can be find in  \cite{qtr1,qtr2,qtr3}).

In this paper we investigate the effects of dissipative
decoherence on the accuracy of the quantum sawtooth map
algorithm. The system Hamiltonian of the exact map reads 
\cite{benenti01,benenti02}
\begin{equation}
  H_s(\hat{n}, \theta) = T \hat{n}^2/2 + k V(\theta) \sum_m \delta(t-m) \;\; .
  \label{eq1}
\end{equation}
Here the first term describes free particle rotation on a ring
while the second term gives kicks periodic in time and 
$\hat{n}=-i\partial/\partial \theta$. The kick potential is
$V(\theta)=-(\theta - \pi)^2/2$ for $0 \leq \theta < 2\pi$.
It is periodically repeated for all other $\theta$
so that the  wave function $\psi(\theta)$ satisfies
the periodic boundary condition $\psi(\theta)=\psi(\theta+2\pi)$.
The classical limit corresponds to $T \rightarrow 0$, $k \rightarrow \infty$
with $K=kT = const$. In these notations
the Planck constant is assumed to be $\hbar=1$
while  $T$ plays the role of an effective dimensionless
Planck constant.

The classical dynamics is described by a symplectic area-preserving map
\begin{equation}
\overline{n}={n}+k(\theta-\pi),
\quad
\overline{\theta}=\theta+T\overline{n} \; .
\label{eq2}
\end{equation}
Using the rescaled momentum variable
$p=Tn$ it is easy to see that the dynamics 
depends only on the chaos parameter $K=kT$.
The motion is stable for $-4<K<0$
and completely chaotic for $K<-4$ and $K>0$ (see \cite{benenti01}
and Refs. therein). 
The map (\ref{eq2}) can be studied on
the cylinder ($p\in (-\infty,+\infty)$), which can also be closed
to form a torus of length $2\pi L$, where $L$ is an integer. 

The quantum propagation on one map iteration is described
by a unitary operator $\hat{U}$ acting on the wave function
$\psi$:
\begin{equation}
\overline{\psi}=\hat{U}\psi = \hat{U}_T \hat{U}_k \psi=
e^{-iT\hat{n}^2/2}
e^{-ikV(\theta)}\psi \; .
\label{eq3}
\end{equation}
The quantum evolution is considered on $N$ quantum momentum levels.
For $N=2^{n_q}$ this evolution can be implemented on a quantum
computer with $n_q$ qubits. The quantum algorithm described 
in \cite{benenti01}
performs one iteration of the quantum map (\ref{eq3})
in $n_g=3n_q^2+n_q$ elementary quantum gates.
It essentially uses the quantum Fourier transform
which allows to go from momentum to phase representation
in $n_q (n_q+1)/2$ gates. The rotation of quantum phases
in each representation is performed 
in approximately $n_q^2$ gates.
Here we consider the case of one classical cell 
(torus with $L=1$ when $T=2\pi/N$) \cite{benenti01}
and the case of dynamical localization
with $N$ levels on a torus and $K \sim 1$, $k = K/T \sim 1$
\cite{benenti02}. Here and below the
time $t$ is measured in number of map iterations.

To study the effects of dissipative decoherence 
on the accuracy of the quantum sawtooth algorithm
we follow the approach 
with the amplitude damping channelused in \cite{carlo2}.
The evolution of the density operator $\rho(t)$ of open system
under weak Markovian noises is given by 
the master equation with Lindblad operators $L_m~(m=1, \cdots, n_q)$:
\beqa
\dot{\rho}=-\frac{i}{\hbar}[H_{eff} \rho -\rho H_{eff}^\dagger]
+\sum_m L_m \rho  L^\dagger_m \; ,
\label{eq4}
\eeqa
where the system  Hamiltonian $H_s$ is related to
the  effective Hamiltonian  
 $H_{eff}\equiv H_s-i\hbar/2 \sum_m L^\dagger_m  L_m$
and $m$ marks the qubit number.
In this paper we assume that
the system is coupled to the environment through 
an amplitude damping channel with 
$L_m=\hat{a}_m\sqrt{\Gamma}$, 
where $\hat{a}_m$ is the destruction operator for $m-$th qubit 
and the dimensionless rate $\Gamma$ gives the decay rate for
each qubit per one quantum gate.
The rate $\Gamma$ is the 
same for all qubits.

This evolution of $\rho$ can be  efficiently simulated
by averaging  over the  $M$ quantum trajectories which evolve
according to the following stochastic differential equation
for states $|\psi^\alpha\rangle~(\alpha=1, \cdots , M)$:
\beqa
|d\psi^\alpha\rangle=-iH_s |\psi^\alpha\rangle dt
+\frac{1}{2}\sum_m ( \langle L^\dagger_m L_m\rangle_\psi \;\;\;\;\;\;\;\; \no
 - L_m^\dagger L_m)
|\psi^\alpha\rangle dt+
\sum_m \left (\frac{L_m}{\sqrt{\langle L_m^\dagger L_m\rangle_\psi}}-1\right )|\psi^\alpha\rangle d N_m \; ,
\label{eq5}
\eeqa
where $\langle  \rangle_\psi$ represents an expectation value
on $|\psi^\alpha\rangle$ and $dN_m$ are stochastic differential variables
defined in the same way as in \cite{carlo2} (see Eq.(10) there).
The above equation can be solved  numerically 
by the quantum Monte Carlo (MC) methods
by letting the state $|\psi^\alpha\rangle$
jump to one of $L_m|\psi^\alpha\rangle/|L_m|\psi^\alpha\rangle|$  states
with probability $dp_m\equiv |L_m|\psi^\alpha\rangle|^2 dt$ \cite{carlo2}
 or evolve to
$(1-iH_{eff} dt/\hbar)|\psi^\alpha\rangle/\sqrt{1-\sum_m dp_m}$
with probability $1-\sum_m dp_m$.
Then, the density matrix can be approximately expressed as
\beq
\label{eq6}
\rho(t) \approx \left \langle  |\psi(t)\rangle \langle \psi(t)|\right \rangle_M
= \frac{1}{M}\sum^M_{\alpha=1}|\psi^\alpha(t)\rangle  \langle\psi^\alpha(t)|
\; ,
\eeq
 where $\langle \rangle_M$ represents an ensemble average over $M$ quantum
 trajectories $|\psi^\alpha(t)\rangle$.
Hence, an expectation value of an operator $O$ is given by
$ \langle O \rangle=Tr(O\rho) \approx \langle O\rangle_M$.

\begin{figure}
\vglue +0.5cm
\includegraphics[width=3.0truein]{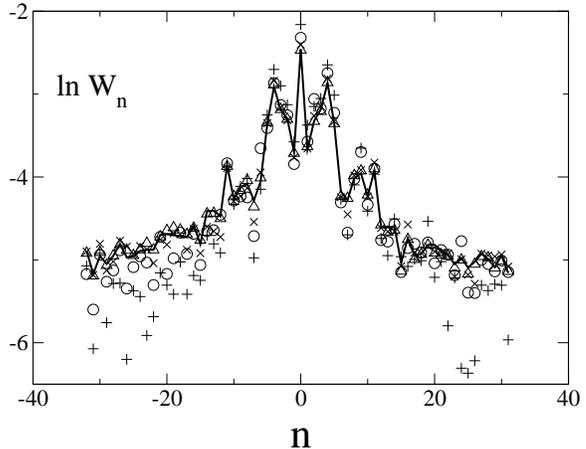}
  \caption{Probability distribution $W_n$ over momentum eigenstates
$n$ in the quantum sawtooth map (\ref{eq3}) at time $t=30$.
The quantum evolution is
simulated by the quantum algorithm with $n_q=6$ qubits
in presence of dissipative decoherence. The dissipation
rate per gate is $\Gamma=0.001$ and the map parameters are
$k=\sqrt{3},K=\sqrt{2}$ with the total number
of states $N=2^{n_q}=64$. The full curve represents the 
 exact solution of the Lindblad equation (\ref{eq4}).
Symbols show the result of quantum trajectories computation
with the number of trajectories
$M=20$(+),
$M=50$(o),
$M=200$(x),
$M=1000$($\triangle$).
The initial state is $n=0$. The logarithm is natural.
  }
  \label{fig1}
\end{figure} 

For the quantum sawtooth algorithm 
the dissipative noise is introduced in
the quantum trajectory context
(Eq. (\ref{eq5})) after each elementary quantum gate
and calculated by  the MC methods.
The same physical process can be described by density matrix theory.
The evolution of density matrix  after single iteration
of quantum sawtooth map
is described by 
\beq
\label{eq7}
\rho'=U_k U_T \rho U_T^\dagger U_k^\dagger \; .
\eeq
 To include the dissipative noise effects
 the density matrix 
further evolves according to 
Eq. (\ref{eq4}) with $H_s=0$
between  consecutive  quantum  gates composing $U_k$ and
 $U_T$.

\begin{figure}
\vglue +0.0cm
  \includegraphics[width=\linewidth]{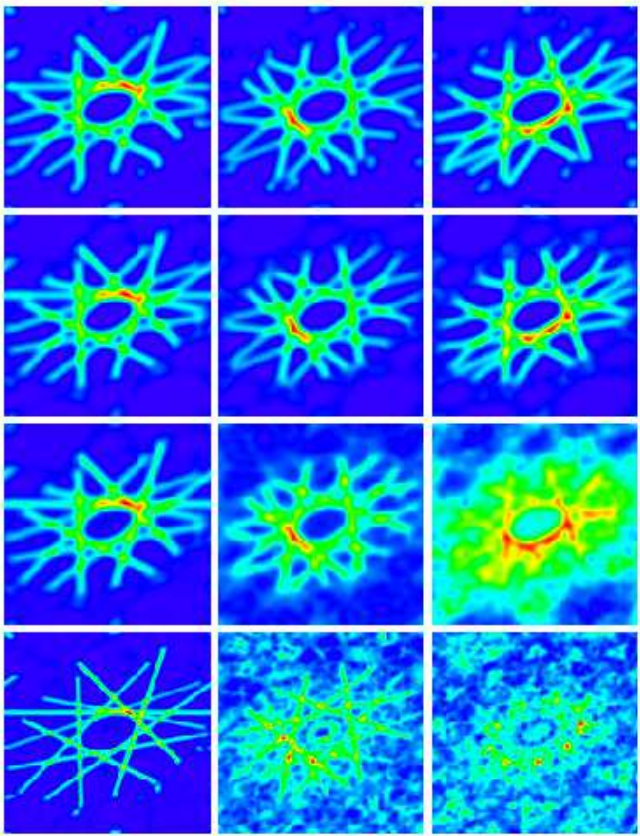}
  \caption{(Color online) 
First top row: classical phase space distribution
obtained from the classical sawtooth map
with a Gaussian averaging over a quantum cell
($N=256$ quantum cells inside whole classical area;
see text). Second row: the corresponding
Husimi function for the quantum sawtooth map
at $n_q=8$ and $\Gamma =0$. Third row:
the Husimi function
obtained with $M=50$ quantum
trajectories in presence of  dissipative decoherence
with rate $\Gamma=0.0005$
and $n_q=8$. Fourth bottom row:
same as for the third row but with
$n_q=10$. Here $K=-0.5$, $T=2\pi/N$
corresponding to $L=1$ and $N=2^{n_q}$
quantum states in the whole classical area.
Columns show distributions 
averaged in the time intervals:
$0 \leq t \leq 9$ (left),
$40 \leq t \leq 49$ (middle),
$90 \leq t \leq 99$ (right).
The initial state is $n \approx 0.1 N$.
Color represents the density from blue/black ($0$)
 to red/gray (maximal value).
  }
  \label{fig2}
\end{figure} 

To test the accuracy of the method of quantum trajectories
we compare its results with the exact 
solution of the Lindblad equation for the density
matrix $\rho$ (\ref{eq4}). The comparison is done for 
the case of dynamical localization of quantum chaos
and is shown in Fig.~1. It shows that 
the dynamical localization is preserved 
at relatively weak dissipation rate $\Gamma$.
It also shows that the quantum trajectories method
correctly reproduces the exact solution of the Lindblad 
equation and that it is sufficient to use $M=50$  trajectories
to reproduce correctly the phenomenon of dynamical localization
in presence of dissipative decoherence of qubits.

\begin{figure}
  \includegraphics[width=\linewidth]{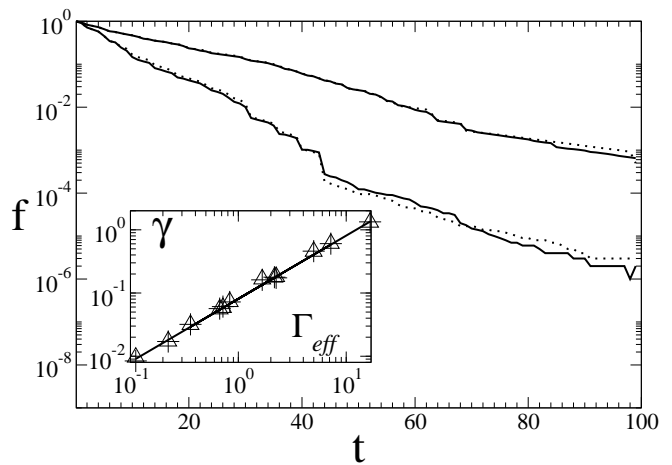}
  \caption{Fidelity $f$ as a function of iteration time $t$.
The upper two curves are for $\Gamma=0.0005$ and
the lower two curves  are for $\Gamma=0.001$.
Here $M=50,n_q=8$,
$k=2^{n_q} K/2 \pi$ and $K=-0.5$ (full curves) 
$K=0.5$ (dotted curves), respectively.
 The initial state is $|n=0\rangle$.
The inset shows  the fidelity decay rate  $\gamma$ 
as a function of
$\Gamma_{eff}\equiv n_q n_g \Gamma $
 with $n_q=4,6,8$.
Here $K=0.5$ (+) and $K=-0.5 (\triangle)$, respectively.
The straight line is the  best fit $\gamma=0.08~\Gamma_{eff}$.
  }
  \label{fig3}
\end{figure} 

To analyze the effects of dissipation rate $\Gamma$
in a more quantitative way we start from
the quasi-integrable case $K=-0.5$ with one classical cell $L=1$
($T=2\pi/N$). 
The classical phase space distribution, averaged over a time interval
and a Gaussian distribution over a quantum cell with an effective
Planck constant, is shown in Fig.~2 
in the first top row (there are $N=2^{n_q}$
quantum cells inside the whole classical phase space).
Such Gaussian averaging of the classical distribution
gives the result which is very close to the
Husimi function in the corresponding quantum case
at $\Gamma=0$ (Fig.~2, second row). We remind that the Husimi
function is obtained by a Gaussian averaging
of the Wigner function over a quantum $\hbar$ cell
(see \cite{husimi} for details). 
In our case the Husimi function
$h(\theta,n)$ is computed through the
wave function of each quantum trajectory
and after that it is averaged over all $M$
trajectories.
The effect of
dissipative decoherence
with $\Gamma=0.0005$ is shown in the third row of Fig.~2.
At $\Gamma=0$ the phase space distribution
remains approximately stationary in time
while for $\Gamma > 0$ it starts to spread 
so that at large times the typical structure
of the classical phase space becomes completely washed out.
This destructive process becomes more rapid with
the increase of the number of qubits even if
$\Gamma$ remains fixed (Fig.~2, fourth bottom row).
One of the reasons is that
$\Gamma$ is defined as a rate per gate 
and the number of gates $n_g = 3n_q^2+n_q$
grows with $n_q$.  However, this is not the 
only reason as it shows the analysis of the fidelity decay.

The fidelity $f$ of quantum algorithm in presence of
dissipative decoherence is defined as 
\begin{equation}
f(t)\equiv \langle \psi_0(t) | \rho(t) |\psi_0(t)\rangle \approx
\frac{1}{M} \sum_\alpha |\langle \psi_0(t) |\psi^\alpha_\Gamma(t)\rangle |^2 
\; ,
\label{eq8}
\end{equation}
where $| \psi_0(t)\rangle $ is the wave function given by the exact
algorithm and $\rho(t)$ is the density matrix of the
quantum computer in presence of decoherence, both are taken
after $t$ map iterations. Here, $\rho(t)$ is expressed 
approximately through the sum over quantum trajectories
(see also \cite{carlo2}).

\begin{figure}
\vglue 0.7cm
  \includegraphics[width=\linewidth]{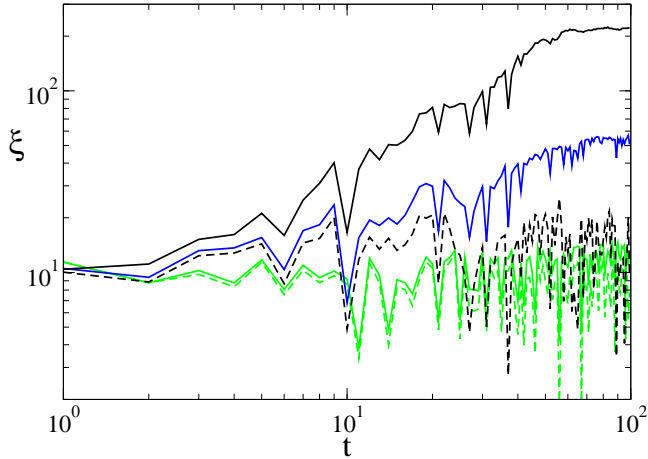}
  \caption{ (Color online) Dependence of the IPR $\xi$  on time $t$
at $\Gamma=0.001$ and $M=50$ 
shown by full curves for $n_q=4$ (green/gray curve),
 $n_q=6$ (blue/black curve), $n_q=8$ (black curve), bottom to top respectively.
The dashed curves show the same cases at  $\Gamma=0$
(bottom dashed curve is for $n_q=4$, top dashed
curve is for $n_q=6$ and $n_q=8$ where $\xi$ values are practically identical).
Here the initial state
is $|n=0\rangle$ and  $k=\sqrt{3}$, $K=\sqrt{2}$.
  }
  \label{fig4}
\end{figure}  

The dependence of fidelity $f(t)$ on time $t$ is shown in Fig.~\ref{fig3}.
At relatively short time $t<50$ the decay is approximately
exponential $f(t) \approx \exp(-\gamma t)$. The decay rate
$\gamma$ is described by the relation
\begin{equation}
\gamma = C \Gamma_{eff} = C n_q n_g \Gamma \; ,
\label{eq9}
\end{equation}
where $C=0.08$ is a numerical constant (see Fig.~\ref{fig3} inset).
The important result of Fig.~\ref{fig3} is that 
the decay of $f(t)$ is not very sensitive
to the map parameters. Indeed, it is not affected by a change of $K$
which significantly modify the classical dynamics
which is quasi-integrable at $K=-0.5$ and
fully chaotic at $K=0.5$.
Another important result is that up to a numerical constant
the relation (\ref{eq9}) follows the dependence found
in \cite{carlo2} for the quantum baker map. This shows that the 
dependence (\ref{eq9}) is really universal. Its physical origin
is rather simple. After one gate the probability of a qubit to stay in 
upper state drops by a factor $\exp(-\Gamma)$ for each qubit (we remind
that $\Gamma$ is defined as a  per gate decay rate).
The wave function of the total system is given by a product
of wave functions of individual qubits
that leads to the fidelity drop by a factor
$\exp( - C n_q \Gamma)$ after one gate and $\exp( - C n_q n_g \Gamma)$
after $n_g$ gates leading to the relation (\ref{eq9}).
In principle, one may expect that the decay of $f(t)$
is sensitive to a number of qubit up states 
in a given wave function since there is no decay
for qubit down states. However, in a context of
a concrete algorithm this number varies in time
and only its average value contributes to the global
fidelity decay. 

The result (\ref{eq9}) gives
the time scale $t_f$ of reliable quantum computation in presence
of dissipative decoherence. On this scale the fidelity should be close
to unity (e.g. $f=0.9$) that gives 
\begin{equation}
t_f \approx 1/(n_q n_g \Gamma) \; ; \;\; N_g = 1/(n_q \Gamma) \; .
\label{eq10}
\end{equation}
Here $N_g=n_g t$ is the total number of quantum gates which can
be performed with high fidelity ($f>0.9$) at given
$n_q$ and $\Gamma$. The comparison with the results
obtained for static imperfections \cite{frahm} of strength $\epsilon$
shows that for them $N_g$ drops more rapidly with $n_q$:
$N_g \sim 1/(\epsilon^2 n_q n_g)$.
Therefore the static  imperfections
destroy the accuracy of quantum computation in a more rapid
way compared to dissipative decoherence.

\begin{figure}
\vglue 0.7cm
  \includegraphics[width=\linewidth]{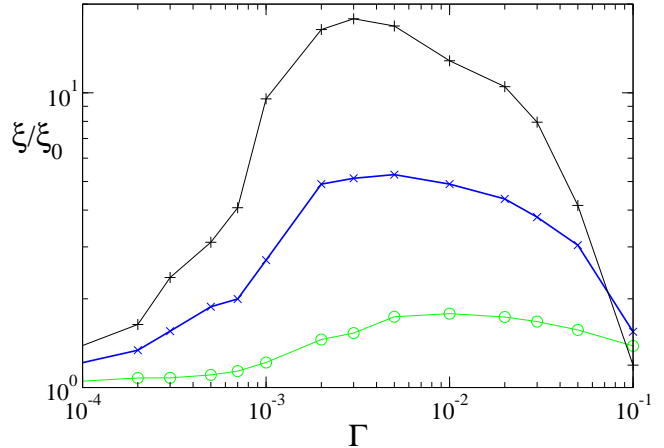}
\vglue -0.2cm
  \caption{(Color online) 
Dependence of the ratio of $\xi$ to its value
$\xi_0$ in the ideal algorithm on the dissipative
decoherence rate $\Gamma$
for $n_q=4$ (green/gray o), $n_q=6$ (blue/black x), $n_q=8$(black +)
from bottom to top. The values $\xi$ and $\xi_0$
are averaged in the time interval $30 \leq t \leq 40$.
Other parameters are as in Fig.~\ref{fig4}.
}
\label{fig5}
\end{figure} 

It is also interesting to analyze the effects of dissipative
decoherence on the dynamical localization. For that, in addition to the 
probability distribution as in Fig.~\ref{fig1},
it is convenient to use the inverse participation ratio (IPR) 
defined as 
$\xi=1/\sum_n |\psi_n|^4
\simeq 1/\sum_n |\langle |\psi_n|^2\rangle_M|^2$
where $\langle|\cdots|\rangle_M$ notes the average over
$M$ quantum trajectories.
This quantity is often used in the problems
with localized wave functions. In fact $\xi$ gives an effective
number of states over which the total probability is distributed. 
The dependence of $\xi$ on time $t$ is shown in Fig.~\ref{fig4}.
It shows that in presence of dissipative decoherence
the dynamical localization is destroyed. Indeed, at large $t$
the value of $\xi$ grows with $n_q$ while for 
the ideal algorithm it is independent of $n_q$.
The physical meaning of this effect is rather clear.
As in Fig.~\ref{fig2} the dissipative decoherence introduces 
some noise which destroys localization.

\begin{figure}
\vglue 0.7cm
  \includegraphics[width=\linewidth]{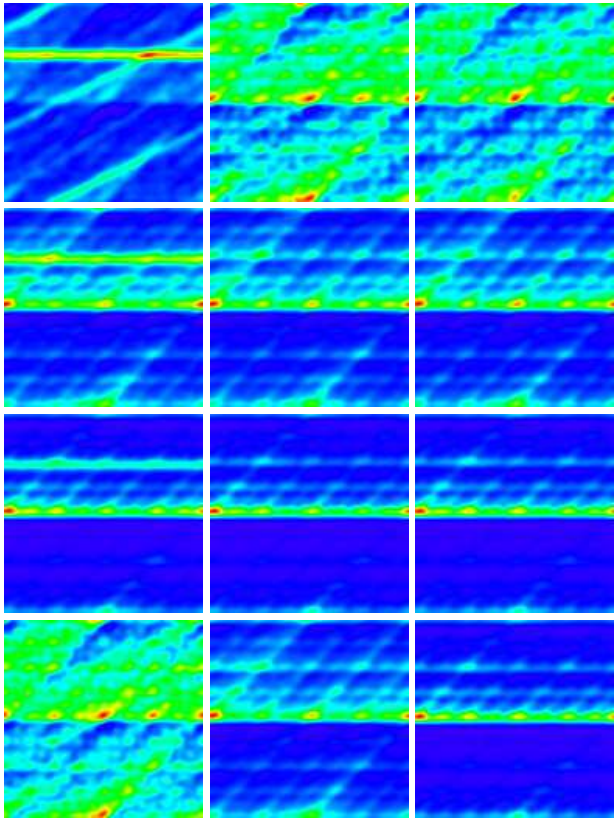}
\vglue -0.2cm
  \caption{(Color online) Each panel shows the 
Husimi distribution for the quantum sawtooth map
algorithm with $K=1$, $T=2\pi/2^{n_q}$ and $n_q=8$.
The top three rows show the cases with the rate
$\Gamma=0.01$,
$\Gamma=0.05$, and $\Gamma=0.1$,
respectively from top to third row.
The initial state is $|n=60 \rangle$
 and $M=50$. The distribution is averaged
in the time interval $0 \leq t \leq 9$ (left column),
$40 \leq t \leq 49$ (middle column),
$90 \leq t \leq 99$ (right column).
The fourth bottom row shows the distribution
for another initial state $|n=0 \rangle$
averaged in the time interval
$90 \leq t \leq 99$ for
$\Gamma=0.01$ (left panel),
$\Gamma=0.05$ (middle panel) , and 
$\Gamma=0.1$ (right panel) (compare with the 
right column of top three rows).
Color represents the density from blue/black ($0$)
to red/gray (maximal value).
}
\label{fig6}
\end{figure}

However, there is also another effect which 
becomes visible at relatively large $\Gamma$. It is shown in Fig.~\ref{fig5}
which gives the ratio of $\xi$ to its value $\xi_0$ in the ideal algorithm.
Thus, at small $\Gamma$ the ration $\xi/\xi_0$ grows with the increase
of $\Gamma$ while it stars to drop
at large $\Gamma$. This  is a manifestation of the
fact that in absence of algorithm the dissipation drives the quantum register
to the state $|n=0\rangle$ with all qubits in down state. 
Even in presence of the quantum algorithm this dissipative effect
becomes dominant at large $\Gamma$ leading to a decrease of 
the ratio $\xi/\xi_0$.

The dissipative effect of decoherence at large values of $\Gamma$
is also clearly seen in the case of quantum chaos ergodic in one
classical cell ($L=1$). At large $\Gamma$ the Husimi distribution
relaxes to the stationary state induced by dissipation
$|n=0 \rangle$ (third row in Fig.~\ref{fig6} at $\Gamma=0.1$).
In a sense this corresponds to a simple attractor in 
the phase space.
The stationary state becomes more complicated with a 
decrease of $\Gamma$ (second row in Fig.~\ref{fig6} at $\Gamma=0.05$).
And at even smaller $\Gamma=0.01$ the stationary  state
shows a complex structure in the phase space
(top first row in Fig.~\ref{fig6}).
It is important to stress that this structure is
independent of the initial state (bottom row in Fig.~\ref{fig6}).
In this sense we may say that 
in such a case the dissipative decoherence
leads to appearance of a quantum strange attractor
in the quantum algorithm. Of course, this stationary
quantum attractor state is very different from the
Husimi distribution generated by the ideal quantum
algorithm. However, it may be of certain interest
to use the dissipative decoherence in quantum algorithms
for investigation of quantum strange attractors
which have been discussed in the context of quantum 
chaos and dissipation (see e.g. Refs.~\cite{graham,maspero,ratchet}).
At the same time we should note that 
the dissipation induced by decoherence 
acts during each gate that makes 
its  effect rather nontrivial due to change of representations
in the map (\ref{eq3}). 

In conclusion, our studies determine the fidelity decay law
in presence of dissipative decoherence
which is in agreement with the results obtained in \cite{carlo2}
for a very different quantum algorithm.
This confirms the universal nature of the established fidelity decay law.
These studies also show that at moderate strength the dissipative
decoherence destroys dynamical localization while 
a strong dissipation leads to localization and 
appearance complex or simple attractor. The effects of dissipative
decoherence are compared with the effects of static imperfections
and it is shown that in absence of
quantum error correction the later give more restrictions on the
accuracy of quantum computations with a large
number of qubits.

This work was supported in part by the EU IST-FET project  EDIQIP.


\begin{thebibliography}{99}
\bibitem{chuang00}  M.A.~Nielsen and I.L.~Chuang {\it Quantum Computation and 
       Quantum Information}, Cambridge Univ. Press, Cambridge (2000).
\bibitem{zurek} W.H.~Zurek, Rev. Mod. Phys. {\bf 75}, 715 (2003).
\bibitem{georgeot00} B.~Georgeot, and D.~L.~Shepelyansky,
       Phys. Rev. E {\bf 62}, 3504 (2000); {\bf ibid.} {\bf 62}, 6366 (2000).
\bibitem{dlsnobel} D.~L.~Shepelyansky, Physica Scripta {\bf T90}, 112 (2001).
\bibitem{cirac} J.I. Cirac and P. Zoller, Phys. Rev. Lett. {\bf 74},
       4091 (1995).
\bibitem{paz1} C.~Miquel, J.~P.~Paz and R.~Perazzo, 
       Phys. Rev. A {\bf 54}, 2605 (1996).
\bibitem{paz2} C.~Miquel, J.P.~Paz and W.H.~Zurek, Phys. Rev. Lett.
        {\bf 78}, 3971 (1997).
\bibitem{song1} P.H.Song and D.L.Shepelyansky, Phys. Rev. Lett. {\bf 86}, 
       2162 (2001).
\bibitem{acat}  B.~Georgeot and D.L.~Shepelyansky,
       Phys. Rev. Lett. {\bf 86}, 5393 (2001); {\bf ibid.}
       {\bf 88}, 219802 (2002).
\bibitem{qcat} A.D.~Chepelianskii, and D.L.~Shepelyansky, Phys. Rev. A
       {\bf 66}, 054301  (2002).
\bibitem{braun} D.~Braun, Phys. Rev. A {\bf 65}, 042317 (2002).
\bibitem{terraneo1} M.~Terraneo, B.~Georgeot and  D.L.~Shepelyansky,
       Eur. Phys. J. D {\bf 22}, 127 (2003).
\bibitem{song2} P.H. Song and I. Kim, Eur. Phys. J. D {\bf 23}, 299 (2003).
\bibitem{levi1} B.~L\'evi, B.~Georgeot and D.L.~Shepelyansky,
       Phys. Rev. E {\bf 67}, 046220 (2003).
\bibitem{bettelli1} S.~Bettelli and D.L.~Shepelyansky,
       Phys. Rev. A {\bf 67}, 054303 (2003).
\bibitem{bettelli2} S.~Bettelli, Phys. Rev. A {\bf 69}, 042310 (2004).
\bibitem{benenti01} G.~Benenti, G.~Casati, S.~Montangero and 
       D.~L.~Shepelyansky, Phys. Rev. Lett. {\bf 87}, 227901 (2001).
\bibitem{terraneo2} M.~Terraneo and D.L.~Shepelyansky, 
       Phys. Rev. Lett. {\bf 90}, 257902 (2003).
\bibitem{benenti02} G.~Benenti, G.~Casati, S.~Montangero and 
       D.~L.~Shepelyansky, Phys. Rev. A {\bf 67}, 52312 (2003).
\bibitem{pomeransky} A.A. Pomeransky, D.L. Shepelyansky, 
       Phys. Rev. A {\bf 69}, 014302 (2004).
\bibitem{frahm} K.~Frahm, R.Fleckinger and D.L.Shepelyansky,
       Eur. Phys. J. D {\bf 29}, 139 (2004).
\bibitem{zhirov} A.A. Pomeransky, O.V.Zhirov and D.L. Shepelaynsky,
       Eur. Phys. J. D {\bf 31}, 131 (2004).
\bibitem{levi2} B.~L\'evi and B.~Georgeot, 
        Phys. Rev. E {\bf 70}, 056218  (2004).
\bibitem{chirikov} B.~V.~Chirikov, in {\it Chaos and Quantum Physics}, 
        Les Houches Lecture Series  {\bf 52}
        Eds. M.-J. Giannoni, A.Voros, and J. Zinn-Justin (North-Holland,
        Amsterdam, 1991), p.443; F.~M.~Izrailev, Phys. Rep. {\bf 196},
        299 (1990).
\bibitem{schack} R.~Schack, Phys. Rev. A {\bf 57}, 1634 (1998).
\bibitem{kr} B.~Georgeot, and D.L.~Shepelyansky, Phys. Rev. Lett. 
       {\bf 86}, 2890 (2001).
\bibitem{cory} Y.S.~Weinstein, S.~Lloyd, J.~Emerson, and D.G.~Cory,
       Phys. Rev. Lett. {\bf 89}, 157902 (2002).
\bibitem{carlo1} G.G.~Carlo, G.~Benenti and G.~Casati, Phys. Rev. Lett.
       {\bf 91}, 257903 (2003).
\bibitem{carlo2} G.G.~Carlo, G.~Benenti, G.~Casati
       and C.~Mejia-Monasterio, Phys. Rev. A
       {\bf 69}, 062317 (2004).
\bibitem{lindblad} G.~Lindblad, Commun. Math. Phys. {\bf 48}, 119 (1976);
       Gorini, A. Kossakowski, and E.C.G. Sudarshan, 
       J. Math. Phys. {\bf 17}, 821 (1976).
\bibitem{qt1}  H.J.~Carmichel, {\it An Open Systems Approach to
        Quantum Optics} (Springer, Berlin, 1993).
\bibitem{qt2} R.~Dum, P.~Zoller, and H.~Ritsch,
        Phys. Rev. A {\bf 45}, 4879 (1992).
\bibitem{qt3} J.~Dalibard, Y.~Castin, and K.~M\o lmer,
        Phys. Rev. Lett. {\bf 68}, 580 (1992).
\bibitem{qt4} N.~Gisin, Phys. Rev. Lett. {\bf 52}, 1657 (1984).
\bibitem{qtr1} R.~Schack, T.A.~Brun, and I.C.~Percival,
        J. Phys. A {\bf 28}, 5401 (1995).
\bibitem{qtr2} A.~Barenco, T.A.~Brun, R.~Schack, and T.P.~Spiller,
        Phys. Rev. A {\bf 56}, 1177 (1997).
\bibitem{qtr3}  T.A.~Brun, Am. J. Phys. {\bf 70}, 719 (2002);
       quant-ph/0301046.
\bibitem{husimi} S.-J. Chang and K.-J. Shi, Phys. Rev. A 
            {\bf 34}, 7 (1986).
\bibitem{graham}T.~Dittrich and R.~Graham,
        Annals of Physics {\bf 200}, 363 (1990).
\bibitem{maspero}G.~Casati, G.~Maspero and D.L.~Shepelyansky, 
        Physica D {\bf 131}, 311 (1999).
\bibitem{ratchet} G.~Carlo, G.~Benenti, G.~Casati and
         D.L.~Shepelyansky, cond-mat/0407702.



\end{thebibliography}
\end{document}